\documentclass[
    aps,
    prb,
    final,
    twocolumn,
    floats,
    floatfix,
    nofootinbib,
    10pt,
]{revtex4-2}
\usepackage[dvipsnames]{xcolor}
\usepackage[
    pdftex,
	colorlinks=true,
	citecolor=blue,
	linkcolor=blue,
	filecolor=magenta,
	urlcolor=blue,
]{hyperref}

\usepackage{graphicx}

\usepackage{amsmath}
\usepackage{bm}
\usepackage{ocgx2}
\usepackage{mathrsfs}
\usepackage{comment}
\usepackage{lipsum}
\usepackage{amsfonts}
\usepackage{amssymb}
\usepackage{amstext}
\usepackage{helvet}
\usepackage{microtype}
\usepackage{esint}
\usepackage{tikz}
\usepackage{mathtools}
\usepackage{scrextend}  
\usepackage{etoolbox}
\usepackage[export]{adjustbox}
\usepackage{bbm}
\usepackage[caption=false]{subfig}
\usepackage{svgcolor}

\newcommand{\iu}{\mathrm{i}\mkern1mu}
\newcommand{\eu}{\mathrm{e}\mkern1mu}

\newcommand*\dd{\mathop{}\!\mathrm{d}}

\definecolor{myred}{HTML}{ff7676}
\definecolor{myblue}{HTML}{2040d2}

\newcommand{\qqoute}[1]{``#1''}

\usetikzlibrary{decorations.pathmorphing,patterns}
\usetikzlibrary{decorations.markings}
\usetikzlibrary{calc}

\newcommand{\comma}{{,\ \ \ }}

\newcommand{\bra}[1]{\left\langle#1\right|}
\newcommand{\ket}[1]{\left|#1\right\rangle}
\DeclarePairedDelimiterX\braket[2]{\langle}{\rangle}{#1 \delimsize\vert #2}

\begin{document}

\title{Transfer Matrix Approach for Topological Edge States}

\author{Rickson Wielian}
\author{Ivan Toftul}
\email{ivan.toftul@anu.edu.au}
\author{Yuri Kivshar}
\email{yuri.kivshar@anu.edu.au}
\affiliation{Nonlinear Physics Center, Research School of Physics, Australian National University, Canberra ACT 2601, Australia}

\date{\today}

\begin{abstract}

We suggest and develop a novel approach for describing topological properties of a periodic system purely from the transfer matrix associated to a unit cell. Our approach uses the Iwasawa decomposition to parametrise the transfer matrix uniquely in terms of three real numbers.
This allows us to obtain simple conditions for the existence of topologically protected edge states and to provide a visual illustration of all possible solutions.
In order to demonstrate our method in action, we apply it to study some generalisations of the Su-Schrieffer-Heeger (SSH) model, such as the tetramer SSH4 model and a dimerised one-dimensional photonic crystal. 
Finally, we also obtained a simple pictorial proof of the Zak phase bulk-edge correspondence for any one dimensional system using this approach.

\end{abstract}

\maketitle

\section{Introduction}

Topological insulators are commonly associated with robust states, localised at the edges of the sample, and which existence are entirely predicted by the topology of the bulk \cite{Lu2014}. As expected, due to their promised robustness, topological edge states currently form an exciting playground for technological applications. For example, topological insulators have been considered as candidate materials to realise qubits for quantum computing \cite{Sarma2015Oct,Blanco}, as well as efficient lasing and harmonic generation in photonic systems \cite{Pilozzi2016May, toftul,Kang2023Jun}.

Of these topological insulators, one of the simplest toy models is the Su-Schrieffer-Heeger (SSH) model, which was first introduced to model electron hopping in long-chain polyenes \cite{Su1981Mar}. Despite its simplicity, the SSH model and its generalisations are often used in experiment to demonstrate techniques for detecting topological phase transitions \cite{Cardano2017Jun}, as well as to demonstrate practical applications of topological edge states \cite{Han2019}. Theoretically, the SSH model and its generalisations serve as a starting point to test new methods for analysing topological insulators \cite{DiLiberto2016Dec,Lieu2018Jan,Arkinstall2017Apr}. Generalisations of the SSH model often also admit exciting nontrivial behaviour, which may shed light on the nature of topological phases \cite{Liu2017Feb}, or present opportunities for technological applications \cite{Bao2023Jan}. On the other hand, due to its simplicity, many models in different fields reduce to the SSH model. For example, the SSH model has recently been realised in acoustic \cite{Coutant2021Jun} and magnonic~\cite{Li2021Jan} systems. 

In this article, we introduce a new method to analyse one-dimensional topological insulators based on top of the well-established transfer matrix approach~\cite{ivchenko2005optical,orfanidis,mackay2022transfer}. 
This transfer matrix method offers an approach complementary to the standard Bloch Hamiltonian method \cite{lan,Chiu2016Aug}. As a preview, a system with $N$ sites per unit cell and $\ell$-nearest neighbour hoppings would admit a Bloch Hamiltonian of size $N\times N$, while the transfer matrix would have a size of $2\ell \times 2\ell$. In addition, while the Bloch Hamiltonian works best with periodic boundary conditions, the transfer matrix method can be naturally equipped with open or closed boundary conditions. A consequence of this is that the edge states appear naturally within the transfer matrix method, so that we may characterise the edge states more explicitly. We thus apply the transfer matrix method to inversion-symmetric models and obtain simple conditions for the appearance of their topologically protected edge states. This is done by uniquely parametrising the transfer matrix in terms of three real numbers using the Iwasawa decomposition of $\mathrm{SL}(2;\mathbb{R})$. This parametrisation reveals the submanifolds corresponding to symmetry-protected systems, and the topology of these submanifolds is intimately connected with the edge states of the system.

The paper is organised as follows. In Sec. \ref{sec:ssht}, we begin by reviewing the transfer matrix method and apply it to solve the semi-infinite SSH model. In Sec. \ref{sec:topologicalprotection}, we demonstrate how this the transfer matrix approach can be employed to obtain topological invariants for the Hamiltonian. This method involves assigning to each Hamiltonian a path on a manifold with the endpoints fixed. We then show how our method connects to bulk-band topological invariants such as the Zak phase. In doing so, we obtain a short, but rigorous proof of the bulk-edge correspondence in general one dimensional systems possessing inversion symmetry. Finally, in Sec. \ref{sec:examples} we show how our method can be applied to more complicated models such as the generalized SSH model and a dimerised photonic crystal.
	
\section{Transfer Matrix of the SSH model}\label{sec:ssht}

\begin{figure}
    \includegraphics[width=\linewidth]{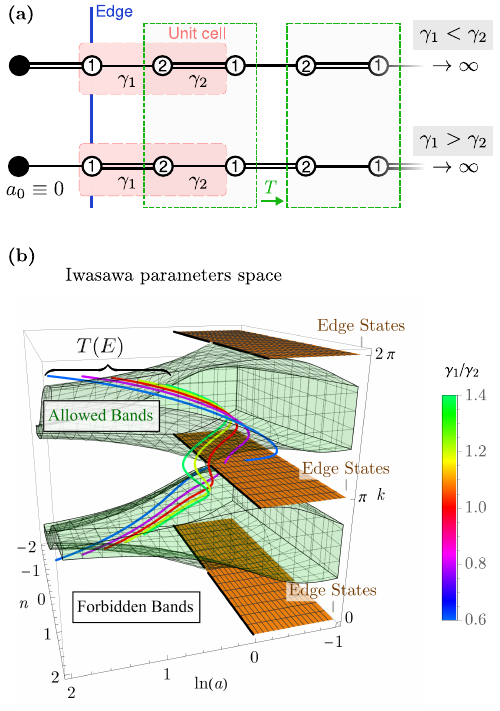}
    \caption{
    \textbf{(a)} Illustration of the SSH model. The open circles denote sites at which the electron can sit, and the lines denote the coupling strength between each site. In the SSH model, the couplings alternate between two values $\gamma_1$ and $\gamma_2$.
    \textbf{(b)} The band structure of the SSH model $(\mu_1=\mu_2)$ embedded in $\mathrm{SL}(2,\mathbb{R})$ (3D Iwasawa parameters space). The parameter $k$ is only defined modulo $2\pi$. The plot is then unfolded in the same way we conventionally unfold band diagrams beyond the first Brillouin zone. Each curve represents a Hamiltonian, or equivalently, a family of $T$-matrices $T(E)$ for a different value of $\gamma_1/\gamma_2$. When the curve touches one of the orange surfaces, an edge state is supported. The decay length of this edge state is given by $|\ln(a)|^{-1}$. The green regions are the allowed bands: Each point on the curve inside the green region corresponds to a bulk propagating mode. The thick black lines on the boundary of the orange surface are the points we want to exclude so that our model to be topologically nontrivial.}
    \label{fig:sshkan}
\end{figure}

We will begin with the most general 1D Hamiltonian supporting two sites per unit cell (Fig.~\ref{fig:sshkan}~(a)),
\begin{align}\label{eqn:ssh}
H =& \sum \limits_{m} \gamma_1 \ket{m,2} \bra{m,1}+ \gamma_2 \ket{m+1,1} \bra{m,2} + \mathrm{h.c.} \nonumber\\
&+v_1\ket{m,1} \bra{m,1}+v_2\ket{m,2} \bra{m,2} ,
\end{align}
where $1$ and $2$ label the two sites in each unit cell. The top line of Eq.~\eqref{eqn:ssh} is the SSH model as originally described in \cite{Su1981Mar}. At this point we have also added to the model alternating on-site potentials for completeness (second line of Eq.~\eqref{eqn:ssh}). This will be important when we generalise to the SSH4 model.

The transfer ($T$) matrix is a powerful method of solving linear one-dimensional systems. It has previously also been considered in the study of topological insulators \cite{Hatsugai1993Oct, GranadaE.J2017Jun, Kunst2019Jun, Dwivedi2016Apr,Arkinstall2017Apr}, though our approach is considerably different. To derive the transfer matrix, it is convenient to rewrite the time-independent Schr\"odinger equation $H\psi=E\psi$ in matrix form:
\begin{align}\label{eqn:masschainxmatrix}
    \begin{pmatrix}
    \mu_1 & \gamma_1 &  &  & & \\
    \gamma_1 & \mu_2 & \gamma_2 &  &  & \\
    & \gamma_2 & \mu_1 & \gamma_1 & & \\
    &  & \gamma_1 & \ddots & \ddots & \\
    &  &  & \ddots & \ddots & \gamma_1 \\
    & & & & \gamma_1 & \mu_2
    \end{pmatrix} 
    \begin{pmatrix}
    a_1 \\ a_2 \\ a_3 \\ \vdots \\ \vdots \\ a_{2N}
    \end{pmatrix}=0,
\end{align}
where we have defined $\mu_i = v_i -E$ and $a_1, a_2, \dots,a_{2N}$ are the amplitudes in each site.

Now, we will attempt to exploit the periodicity of the matrix in Eq.~\eqref{eqn:masschainxmatrix}, and we'll see that this naturally leads us to a transfer matrix for the SSH model.

Solving the matrix equation \eqref{eqn:masschainxmatrix} within first unit cell
\begin{equation}
\begin{pmatrix}
\gamma_2 & \mu_1 & \gamma_1 & 0\\
0 & \gamma_1 &\mu_2 & \gamma_2
\end{pmatrix} \begin{pmatrix}
    a_0\\a_1\\a_2\\a_3
\end{pmatrix}=0 
\end{equation}
using Gauss-Jordan elimination, we obtain the reduced row echelon form which can be written in the block form:
\begin{equation}
    \begin{pmatrix}
        -T & \mathbbm{1}\end{pmatrix} \begin{pmatrix}
    a_0\\a_1\\a_2\\a_3
\end{pmatrix}=0 
    , \quad T = \begin{pmatrix}
-\frac{\gamma_2}{\gamma_1} & -\frac{\mu_1}{\gamma_1} \\
\frac{\mu_2}{\gamma_1} &  \frac{\mu_1 \mu_2}{\gamma_1 \gamma_2} - \frac{\gamma_1}{\gamma_2}
\end{pmatrix},
\label{eqn:massunitcell}
\end{equation}
where $\mathbbm{1}$ is a $2 \times 2$ identity matrix. The condition that a solution is in the null-space of this reduced matrix becomes exactly the definition of the transfer matrix: 
\begin{equation}
    \bm{a}_{i+1}=T(E) \bm{a}_i.
\end{equation}
Here, the dependence of $T$ on $E$ comes in through the energy-dependence of $\mu_1,\mu_2$ in Eq.~\eqref{eqn:massunitcell}. Also, we define $\bm{a}_i:=(a_{2i},a_{2i+1})^T$ where $a_{2i}$ is the rightmost amplitude of the previous unit cell and $a_{2i+1}$ is the first (leftmost) amplitude of the current unit cell. In this way, for a system of $N$ sites, we have introduced 2 more amplitudes in our transfer matrix description, $a_0$ and $a_{N+1}$. These will be important in the next paragraph when we consider boundary conditions. The gaussian elimination process also shows that in general, any one-dimensional periodic Hamiltonian admits a set of transfer matrices that is unique up to basis changes or symmetry transformations\footnote{This freedom is important as any hermitian tridiagonal Hamiltonian admits a gauge transformation to a real-symmetric tridiagonal Hamiltonian. This choice of basis will become important for the arguments in the later sections.}.  %

The main advantage of the $T$-matrix formalism appears when we consider semi-infinite and finite chains. For the SSH model, the Dirichlet boundary conditions as applied in Eq.~\eqref{eqn:masschainxmatrix} is equivalent to setting the amplitude $a_0$ to the left of the first amplitude to be zero. As a result, the amplitudes over the entire chain are determined by the amplitude at the first site by the $T$-matrix:
\begin{align}
\begin{pmatrix}\label{eqn:sshtmatrix}
a_{2k}\\ a_{2k+1}
\end{pmatrix}=\begin{pmatrix}
-\frac{\gamma _2}{\gamma _1} & -\frac{\mu _1}{\gamma _1} \\
\frac{\mu _2}{\gamma _1} & \frac{\mu _1 \mu _2}{\gamma _1 \gamma _2}-\frac{\gamma _1}{\gamma _2} \\
\end{pmatrix}^k\begin{pmatrix}
0\\ a_1
\end{pmatrix}.
\end{align}
For the remaining of this article, we will call any basis which satisfies the above condition a ``\textit{position basis}''. In the SSH model, this coincides exactly with the conventional meaning of the position basis, {where each basis vector corresponds to a specific site on the lattice}.
When the chain is semi-infinite, this condition, along with the condition that $T^k (^{0} _{1})$ stays bounded in the chain, recovers the exact eigenmodes of the chain.
This includes both the Bloch bands and the edge modes.
It should be noted that the Bloch theorem no longer applies in the semi-infinite case as there is no translation symmetry in the system. 
To recover the exact eigenspectrum one needs to consider each mode as a superposition of counter-propagating waves in the infinite chain which fulfills the condition~\eqref{eqn:sshtmatrix}. 

In general, for a $T$-matrix with $\det T =1$, the above reasoning reduce to simple conditions. 
For the bulk modes, for which Bloch's theorem is applicable, each propagating mode must be an eigenvector of $T$ with eigenvalues $\eu^{\pm \iu K}$, where $K$ is the system quasi-momentum. Since the trace of the matrix is equal to the sum of its eigen values, this equivalent to $\operatorname{Tr}(T) = 2 \cos(K)$ and the above mentioned condition simplifies to
\begin{equation}
    |\mathrm{Tr}(T)|\leq 2 \quad \implies\quad  \text{bulk mode}.
\end{equation}
For an edge mode, since the boundary condition imposes $a_0=0$, the $T$-matrix must instead satisfy
\begin{equation}
\label{eq:edge_mode_condition}
    T\begin{pmatrix}
             0\\1
         \end{pmatrix} = \alpha \begin{pmatrix}
             0\\1
         \end{pmatrix} \comma \alpha \in (-1, 1)
 \implies  \text{edge mode}.
\end{equation}
The condition $|\alpha|<1$ ensures the mode is localized at the edge.
Notably, when Eq.~\eqref{eq:edge_mode_condition} is satisfied, we automatically have $|\mathrm{Tr}(T)| >2,$ so a $T$ matrix cannot support both an edge mode and a bulk mode at the same energy $E$.

The only cases where Bloch's theorem still applies are the cases when both eigenvalues of $T$ are equal, and edge modes, where the left and right edge modes do not couple. So, the left edge mode is obtained by the choice of $E$ which makes $(0,1)^{T}$ an eigenvector of the $T$ matrix. This is easily read off from \eqref{eqn:sshtmatrix} to be $\mu_1=v_1-E=0$. Finally, substituting this into our $T$-matrix, we recover the well-known condition that the left edge mode exists if and only if $|\gamma_1/\gamma_2|<1$. A main advantage of this technique is that we are able to obtain the condition for an edge mode in the SSH model without any symmetry or hermiticity requirement on the Hamiltonian. As such, the condition $|\gamma_1/\gamma_2|<1$ works just as well for $\gamma_1,\gamma_2\in\mathbb{C}$ (non-hermitian models, e.g. systems with gain, nonreciprocal hopping, etc), as well as for symmetry-broken SSH models with jagged on-site potentials.

\section{Topological Protection in the Transfer Matrix Formalism}\label{sec:topologicalprotection}

\subsection{The SL(2,$\mathbb{R}$) Picture for Topological Phases}

The general property that the $T$-matrix always has $\det{T}=1$ implies that $T$ is an element of the special linear group $\mathrm{SL}(2;\mathbb{R}\text{ or }\mathbb{C})$. While in principle the $T$-matrix may be complex, for the purposes of this article we will restrict ourselves to the case where $T$ is real for the sake of clarity and simplicity\footnote{The complex case is largely analogous, but presents some unimportant subtleties.}. The fact that $\mathrm{SL}(2,\mathbb{R})$ is a 3-dimensional smooth manifold promises that each $T$ matrix can be plotted as a point in 3D in a well-behaved manner. Therefore, as a Hamiltonian gives rise to a family of $T(E)$ matrices, it is helpful for visualisation purposes, to plot $T(E)$ as a parametric curve in this space. Here $E$ is the scalar appearing in the the eigenvalue equation $(H-E)\psi=0$ as in \eqref{eqn:masschainxmatrix}.

To assign a $T$ matrix a set of three numbers, we use a result from Iwasawa \cite{Iwasawa1949} which states that any matrix in $\mathrm{SL}(2,\mathbb{R})$ can be written uniquely as a product of three matrices of the form
\begin{align}\label{eqn:kandecomposition}
T = \begin{pmatrix}
\cos(k) & \sin(k)\\
-\sin(k) & \cos(k)
\end{pmatrix}
\begin{pmatrix}
a^{-1} & 0\\
0 & a
\end{pmatrix}
\begin{pmatrix}
1&0\\
n&1
\end{pmatrix},
\end{align}
where $k,n\in \mathbb{R},a>0$ are the \textit{Iwasawa parameters}, and they are smooth functions of the entries $T_{ij}$. 
By doing so, we can parameterise $T(E)$ as a curve in a three-dimensional space, where the coordinates are taken to be the Iwasawa parameters: $k(E)$, $a(E)$, $n(E)$. We plot a family of curves with various choices of $\gamma$ in Fig.~\ref{fig:sshkan} for the SSH model. Since $\det T \in\mathrm{SL}(2,\mathbb{R})$ the condition that the bulk states must stay bounded in the chain is automatically satisfied and we may identify the transfer matrices supporting bulk states by simply looking at the condition $\mathrm{Tr}(T)\leq 2$. On the other hand for a $T$ matrix to support an edge state we follow condition imposed by Eq.~\eqref{eq:edge_mode_condition}: we require $(0,1)$ to be an eigenvector of $T$ with eigenvalue $\alpha$ less than one, $|\alpha|< 1$. In terms of the Iwasawa parameters, this imposes $n=0$, $k\in\pi\mathbb{Z}$, and $a<1$. This way, $\alpha=\cos (k) a=\pm a$.

Fig.~\ref{fig:sshkan} shows that the bands touch only at the points $k\in\pi\mathbb{Z},\ln(a)=n=0$. When $\ln(a)=0,n\neq 0$, only one of the lower or upper green regions touch the edge state surface. This means that edge states can disappear without the need for the bands to touch. 

Topological insulators are characterised by edge states which cannot disappear unless the material passes through a conducting state~\cite{Qi2011Oct}. In the above picture, this means that we need to somehow exclude all the points $k=\pi\mathbb{Z},\ln(a)=0,n\neq 0$ (these are the thick black lines in \ref{fig:sshkan}). Any kind of constraint on the Hamiltonian or $T$-matrix which excludes these points will do the job just fine, but this is usually done using symmetries. In agreement with Zak's argument~\cite{zak}, parity symmetry automatically gives us the following relation for the Iwasawa parameters
\begin{align}\label{eqn:tinversion}
\sigma_x T\sigma_x = T^{-1} \quad  \implies \quad  n=\left(\frac{1}{a^2}-1\right) \tan (k),
\end{align}
where $\sigma_x$ is the first Pauli matrix. So, when $a=1,$ $n$ is automatically zero. This means that if we have two Hamiltonians, $H_1$ and $H_2$, with $H_1$ supporting an edge state and $H_2$ not supporting an edge state, there is no way to continuously deform $H_1$ into $H_2$ without either breaking parity symmetry, or crossing a conducting state at $n=0$.

This way, we may define a topological invariant for the Hamiltonian given by the winding number of $T(E)$ around the points $\ln(a)=0, k\in \pi\mathbb{Z}$.

This shows that we now have a three-step process to find a topological family of Hamiltonians using the transfer matrix.
\begin{enumerate}
    \item Start with a family of Hamiltonians $H$ and obtain the $T$-matrix $T(H)$, either by row reduction or by general arguments \cite{ivchenko2005optical}. 
    \item Find a symmetry $\mathscr{S}$ of the system so that $T(H) = T(\mathscr{S}^{-1}H\mathscr{S})$, and see if this excludes to the pure shears $k\in \pi\mathbb{Z},a=1,n\neq 0$. If not, find the conditions for which this holds.
    \item Apply the conditions to $T$ and see what further conditions are required for it to support edge states. These conditions can be applied to $H$, to obtain a Hamiltonian which supports edge states protected by the symmetry $\mathscr{S}$.
\end{enumerate}

For the family of Hamiltonians \eqref{eqn:ssh} the only constraint put by parity symmetry is that we must have $v_1 = v_2$. This recovers the original SSH model as written down by the original authors \cite{ssh}.

\subsection{Proof of the Zak Phase Bulk-Edge Correspondence}\label{sec:proofzakphase}

Here, we show how our formalism so far is connected to Bloch band topological invariants. In doing so, we will obtain a rigorous but pictorially simple proof of the bulk-edge correspondence. We stress that our results hold for any number of sites per unit cell, as well as for continuous quantum mechanical or photonic systems. This will become more clear in Sec. \ref{sec:ssh4} where we treat the case of $4$ sites per unit cell more explicitly. But for now, we will simply highlight a few key features when we increase the number of sites per unit cell.

For $N$ sites per unit cell, while the Bloch Hamiltonian grows to a size of $N\times N$, as long as interactions are kept to nearest-neighbours, the transfer matrix stays at a size of $2\times 2$. While the Bloch Hamiltonian acts on the amplitudes on the whole unit cell $(a_{jN+1},a_{jN+2},\cdots,a_{(j+1)N}),$ the transfer matrix acts only on the amplitudes $(a_{jN},a_{jN+1})$ to give $(a_{(j+1)N},a_{(j+1)N+1}).$

To connect between the transfer matrix picture and the Bloch Hamiltonian picture, let us first begin by analysing the Bloch eigenvector. Parity and time reversal\footnote{The Hamiltonian considered in this paper has real entries, so it trivially satisfies time-reversal symmetry. This statemenet however is true for PT-symmetric Hamiltonians which are allowed to be complex.} symmetry lets us write any Bloch eigenvector in the position basis \eqref{eqn:sshtmatrix} as 
\begin{align}
\ket{K} = \mathcal{N} \, (1,\cdots,\eu^{\iu \phi(K)}),
\label{eq:bloch}
\end{align}
where $\mathcal{N}$ is a normalisation constant, $\phi(K)$ is a real function, and the entries in the dots are not important (see Appendix \ref{app:zakphaseeigvec} for details). 

Using this fact, we may generally recover the Zak phase \cite{zak} as
\begin{align}\label{eqn:zakphaseeigvec}
    \phi_\text{Zak} \equiv \frac{1}{\iu}\int \limits_0^{2\pi} \bra{K}\partial_K \ket{K} \dd K = \frac{\phi(2\pi) - \phi(0)}{2},
\end{align}
where $\ket{K}$ is the Bloch eigenvector of the system. This result is well-known for the case of 2 sites per unit cell \cite{lan}. We provide details of the general case in Appendix~\ref{app:zakphaseeigvec}.

\begin{figure}
    \centering
    \includegraphics[width=\linewidth]{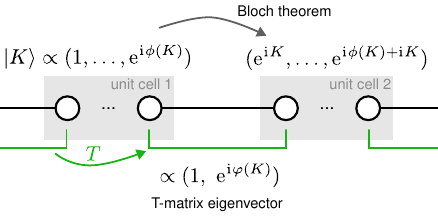}
    \caption{Illustration of the connection between the transfer matrix picture and the Bloch Hamiltonian picture. The $T$-matrix connects only the leftmost and rightmost  nodes of unit cells.}
    \label{fig:phase_derive}
\end{figure}

While this suffices to show that the Zak phase is quantised, we are still required to show the bulk-edge correspondence for our Zak phase. 
To do this, we note that the phase $\phi$ may be obtained from the transfer matrix as 
\begin{align}\label{eqn:phaserelation}
\phi(K) = K-\varphi(K),
\end{align}
where $\varphi(K)$ is the relative phase between the two entries of the $T$-eigenvector with eigenvalue $\eu^{\iu K}$ (Fig.~\ref{fig:phase_derive}). At the same time, this can be interpreted as the relative phase between the last element of the Bloch eigenvector $\ket{K}$ and the first element $\ket{K}\eu^{\iu K}$ of the next unit cell.

It should be noted here that both $\phi$ and $\varphi$ depend on the choice of basis. This is why the middle expression in \eqref{eqn:phaserelation} must be written in a basis-dependent way. This basis dependence will not be a problem though since in the end we are only interested in differences $\phi(2\pi)-\phi(0)$ and $\varphi(2\pi)-\varphi(0)$, where the basis dependence will cancel.

\begin{figure}[t]
    \includegraphics[width=\linewidth]{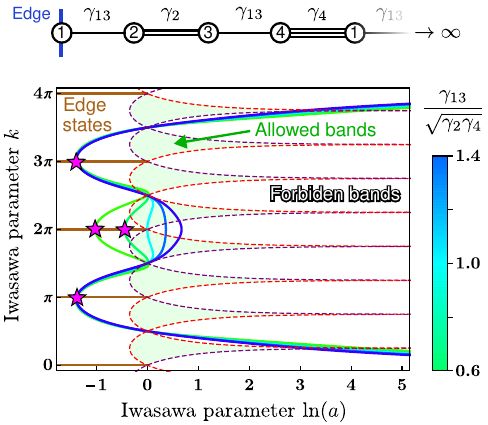}
    \caption{Bulk-edge correspondence for the Zak phase, as viewed in SL(2,$\mathbb{R}$), and projected to $n$ given by Eq.~\eqref{eqn:tinversion}. For clarity we have included example $T(E)$ paths using the SSH4 model described in \eqref{eqn:ssh4t}. Note that the parameter $k$ is only defined modulo $2\pi$. The plot has been unfolded in much the same way as a band diagram can be unfolded beyond the first Brillouin zone. The green regions denote transfer matrices which support bulk modes (allowed bands regions), while the brown lines denote transfer matrices which support edge modes. The purple dashed boundaries denote the points where $\phi=\pi$, and the red dashed curves denote the points where $\phi=0$. The Zak phase $\phi_{\text{Zak}}$ is $\pi$ when the $T(E)$ curve enters and exits from boundaries of different colours, and it is 0 when the two colours are the same. As such, we see that the creation of edge states (magenta stars) is necessarily accompanied by the change in Zak phase of the two neighbouring bands. }
    \label{fig:ssh4kan}
\end{figure}

The parity symmetry condition \eqref{eqn:tinversion} along with the condition 
$|\mathrm{Tr}(T)|=2$,
that the Bloch eigenvector is $K\equiv0 \pmod {\pi}$, fixes $T$ to lie in one of four branches of solutions (See Appendix \ref{app:foursolutions}). As an example, one of the possible solutions is given by
\begin{equation}
    T = \left(
    \begin{array}{cc}
    -\frac{\cos (k)}{\sin (k)+\cos (k)} & \frac{\sin (k)}{\sin (k)+\cos (k)} \\
    -\frac{\sin (k)}{\sin (k)+\cos (k)} & -\frac{2 \sin (k)+\cos (k)}{\sin (k)+\cos (k)} \\
    \end{array}
    \right),
\end{equation}
where $k$ here is the Iwasawa $k$ parameter, with all other solutions differing by signs. Importantly, all four solutions fix $\varphi = 0 \pmod{\pi}$. The results are summarised in Fig.~\ref{fig:ssh4kan}, with the $T(E)$ paths for the SSH4 model included for illustrative purposes. From the figure, we see that the edge states live in the region $\ln(a)<0$. We note that when $k=\pi/2 \pmod{2\pi}$ Eq.~\eqref{eqn:tinversion} dictates that $\ln(a)=0$. As such, each bulk band lives in the interval $k\in (0,\pi)$ modulo $\pi$. Hence, for the transfer matrix curve $T(E)$ to cross between two neighbouring bands, while crossing from $\ln(a)<0$ to $\ln(a)>0$, the Zak phase of the two neighbouring bands must be $\pi$.

This shows that in general the number of edge state in the forbidden zone above the $j^\text{th}$ band is the sum of Zak phases of all the lower bands, modulo 2:
\begin{align}
    n_{\text{edge}}^{(j)} = \frac{1}{\pi} \sum_{\overset{\text{bands}}{i\leq j}} \phi_\text{Zak}^{(i)} \pmod 2.
\end{align}

This form of the bulk-edge correspondence has previously been proven for interfaces between two binary photonic crystals, and numerically confirmed for continuous-index systems \cite{Xiao2014Apr}. However, our result proves this assertion rigorously for any discrete or continuous system possessing inversion symmetry, and works for different interfaces.

\section{Specific examples}\label{sec:examples}

\subsection{Tetramer SSH4 Model}\label{sec:ssh4}
    
Recently there has been considerable interest, in both theory and experiment, in increasing the number of sites per unit cell \cite{Zhang2021Dec,Martinez,Bao2023Jan,Zhou2023Feb,Xie2019May,Marques2019Jul,Marques2020Jan}. In the trimer case, the results are not too interesting. Numerical calculations show that the number of edge states simply correspond to the number of sites to the left of the weakest coupling. The three possible cuts are shown in Fig.~\ref{fig:trimer}, and we see that the number of edge modes agrees with our intuition.
    
\begin{figure}[h]
    \begin{tikzpicture}
    \node[circle,draw=black, fill=white, inner sep=0pt,minimum size=8pt] (m1) at (-4,0) {};
    \node[circle,draw=black, fill=white, inner sep=0pt,minimum size=8pt] (m2) at (-3,0) {};
    \node[circle,draw=black, fill=white, inner sep=0pt,minimum size=8pt] (m3) at (-2,0) {};
    \node[circle,draw=black, fill=white, inner sep=0pt,minimum size=8pt] (m4) at (-1,0) {};
    \node[circle,draw=black, fill=white, inner sep=0pt,minimum size=8pt] (m5) at (0,0) {};
    \node[circle,draw=black, fill=white, inner sep=0pt,minimum size=8pt] (m6) at (1,0) {};
    \node[] (m7) at (2,0) {\dots};
    
    \draw[thick,myblue] ($(m1)-(0.3,0.3)$) rectangle ($(m1)+(0.3,0.5)$) node[midway, above, yshift=0.5cm]{\sffamily Edge};
    \draw[thick,myred] ($(m2)-(0.3,0.3)$) rectangle ($(m4)+(0.3,0.5)$) node[midway, above, yshift=0.5cm]{\sffamily Bulk};
    \draw[thick,myred] ($(m5)-(0.3,0.3)$) rectangle ($(m7)+(0.3,0.5)$) node[midway, above, yshift=0.5cm]{\sffamily Bulk};
    
    \node[circle,draw=black, fill=white, inner sep=0pt,minimum size=8pt] (m11) at (-4,-1) {};
    \node[circle,draw=black, fill=white, inner sep=0pt,minimum size=8pt] (m12) at (-3,-1) {};
    \node[circle,draw=black, fill=white, inner sep=0pt,minimum size=8pt] (m13) at (-2,-1) {};
    \node[circle,draw=black, fill=white, inner sep=0pt,minimum size=8pt] (m14) at (-1,-1) {};
    \node[circle,draw=black, fill=white, inner sep=0pt,minimum size=8pt] (m15) at (0,-1) {};
    \node[circle,draw=black, fill=white, inner sep=0pt,minimum size=8pt] (m16) at (1,-1) {};
    \node[] (m17) at (2,-1) {\dots};
    
    \draw[thick,myblue] ($(m11)-(0.3,0.3)$) rectangle ($(m12)+(0.3,0.5)$);
    \draw[thick,myred] ($(m13)-(0.3,0.3)$) rectangle ($(m15)+(0.3,0.5)$);
    \draw[thick,myred] ($(m16)-(0.3,0.3)$) rectangle ($(m17)+(0.3,0.5)$);
    
    \node[circle,draw=black, fill=white, inner sep=0pt,minimum size=8pt] (m21) at (-4,-2) {};
    \node[circle,draw=black, fill=white, inner sep=0pt,minimum size=8pt] (m22) at (-3,-2) {};
    \node[circle,draw=black, fill=white, inner sep=0pt,minimum size=8pt] (m23) at (-2,-2) {};
    \node[circle,draw=black, fill=white, inner sep=0pt,minimum size=8pt] (m24) at (-1,-2) {};
    \node[circle,draw=black, fill=white, inner sep=0pt,minimum size=8pt] (m25) at (0,-2) {};
    \node[circle,draw=black, fill=white, inner sep=0pt,minimum size=8pt] (m26) at (1,-2) {};
    \node[] (m27) at (2,-2) {\dots};
    
    \draw[thick,myred] ($(m21)-(0.3,0.3)$) rectangle ($(m23)+(0.3,0.5)$);
    \draw[thick,myred] ($(m24)-(0.3,0.3)$) rectangle ($(m26)+(0.2,0.5)$);
    
    \node[] () at ($(m1)+(-0.5,0)$) {1};
    \node[] () at ($(m11)+(-0.5,0)$) {2};
    \node[] () at ($(m21)+(-0.5,0)$) {0};
    
    \draw[thick] (m1)--(m2) node[midway, above]{$\gamma_{1}$};
    \draw[double, thick] (m2)--(m3)node[midway, above]{$\gamma_{2}$};
    \draw[double, thick] (m3)--(m4)node[midway, above]{$\gamma_{3}$};
    \draw[thick] (m4)--(m5) node[midway, above]{$\gamma_{1}$};
    \draw[double, thick] (m5)--(m6)node[midway, above]{$\gamma_{2}$};
    \draw[double, thick] (m6)--(m7)node[midway, above]{$\gamma_{3}$};
    
    \draw[double, thick] (m11)--(m12) node[midway, above]{$\gamma_{1}$};
    \draw[thick] (m12)--(m13)node[midway, above]{$\gamma_{2}$};
    \draw[double, thick] (m13)--(m14)node[midway, above]{$\gamma_{3}$};
    \draw[double, thick] (m14)--(m15) node[midway, above]{$\gamma_{1}$};
    \draw[thick] (m15)--(m16)node[midway, above]{$\gamma_{2}$};
    \draw[double, thick] (m16)--(m17)node[midway, above]{$\gamma_{3}$};
    
    \draw[double, thick] (m21)--(m22) node[midway, above]{$\gamma_{1}$};
    \draw[double, thick] (m22)--(m23)node[midway, above]{$\gamma_{2}$};
    \draw[thick] (m23)--(m24)node[midway, above]{$\gamma_{3}$};
    \draw[double, thick] (m24)--(m25) node[midway, above]{$\gamma_{1}$};
    \draw[double, thick] (m25)--(m26)node[midway, above]{$\gamma_{2}$};
    \draw[thick] (m26)--(m27)node[midway, above]{$\gamma_{3}$};
    
    \end{tikzpicture}
    
    \caption{Summary of the numerical results for the three different cuts in the SSH3 model. We may interpret the bulk modes as modes that arise due to the collective excitation of the red-boxed sites to the right of the weakest coupling (single-lines). The edge modes are then interpreted as the excitation of blue-boxed leftover sites. The number of left edge modes are given on the left, and they correspond to the number of sites to the left of the weakest coupling (modulo 3). This agrees with the grouping argument frequently used for one-dimensional topological systems \cite{Batra}.}
    \label{fig:trimer}
\end{figure}
    
Adding another site per unit cell, numerical calculations for the SSH4 model show that this intuitive picture does not always hold. As such, it may prove useful to investigate analytically what happens when we have $4$ sites per unit cell. Explicitly, this SSH4 model is described by the following Hamiltonian, analogous to equation \eqref{eqn:ssh}:
\begin{align}\label{eqn:ssh4}
H =& \sum \limits_{m} \gamma_1 \ket{m,2} \bra{m,1}+ \gamma_2 \ket{m,3} \bra{m,2} \nonumber\\
&+\gamma_3 \ket{m,4} \bra{m,3}+ \gamma_4 \ket{m+1,1} \bra{m,4} + \mathrm{h.c.} \nonumber\\
&+v_1\ket{m,1} \bra{m,1}+v_2\ket{m,2} \bra{m,2} \nonumber\\
&+v_3 \ket{m,3} \bra{m,4}+v_4\ket{m,4} \bra{m,4} .
\end{align}

The $T$-matrix for this model can be obtained using the same process \eqref{eqn:massunitcell} for each $2\times 2$ half-unit-cell. Then, the $T$-matrix for a full unit cell is given by
\begin{align}
    T=\left(
    \begin{array}{cc}
    -\frac{\gamma _2}{\gamma _3} & -\frac{\mu _3}{\gamma _3} \\
    \frac{\gamma _2 \mu _4}{\gamma _3 \gamma _4} & \frac{\mu _3 \mu _4}{\gamma _3 \gamma _4}-\frac{\gamma _3}{\gamma _4} \\
    \end{array}
    \right)\left(
    \begin{array}{cc}
    -\frac{\gamma _4}{\gamma _1} & -\frac{\mu _1}{\gamma _1} \\
    \frac{\gamma _4 \mu _2}{\gamma _1 \gamma _2} & \frac{\mu _1 \mu _2}{\gamma _1 \gamma _2}-\frac{\gamma _1}{\gamma _2} \\
    \end{array}
    \right).
\end{align}
The left edge state is then given by the condition $T_{12}=0$ and the right edge state is given by the condition $T_{21}=0$. %
Looking at Eq.~\eqref{eqn:tinversion}, we see that for the SSH4 transfer matrix to be parity symmetric for all choices of parameters, it must satisfy the constraints $\gamma_1=\gamma_3 \equiv \gamma_{13}$ and $\mu_4=\mu_1\equiv\bar\mu-\delta$ and $\mu_3=\mu_2\equiv\bar\mu+\delta$. However, we may consider only the case $\delta=0$, and treat $\delta\neq 0$ as a parity symmetric perturbation. As a result, the topological SSH4 transfer matrix is
\begin{align}\label{eqn:ssh4t}
    T=\left(
    \begin{array}{cc}
    \frac{\gamma _4 \left(\gamma _2^2-\bar\mu^2\right)}{\gamma _2 \gamma _{13}^2} & \frac{\bar\mu \left(\gamma _2^2+\gamma _{13}^2-\bar\mu^2\right)}{\gamma _2 \gamma _{13}^2} \\
    \frac{\bar\mu^3-\left(\gamma _2^2+\gamma _{13}^2\right) \bar\mu}{\gamma _2 \gamma _{13}^2} & \frac{\left(\gamma _{13}^2-\bar\mu^2\right){}^2-\gamma _2^2 \bar\mu^2}{\gamma _2 \gamma _4 \gamma _{13}^2} \\
    \end{array}
    \right).
\end{align}
We see that when $\bar\mu=0$, the $T$ matrix may support a topological edge state with $T$-eigenvalue $\gamma_3 \gamma_1 / (\gamma_4\gamma_2)$.
That is, the SSH4 model seems to act just like 2 jagged copies of the dimer SSH model, where the $T$-eigenvalue for the edge state is just the product of the eigenvalues for each dimer. However, we note that, even if $\gamma_1>\gamma_2$, the SSH4 chain may support a topological edge state as long as $\gamma^2_{13}/\gamma_2 < \gamma_4$. Furthermore, even if $\gamma_4$ is the weakest coupling, we can always choose $\gamma_{13},\gamma_2$ so that the eigenvalue remains less than 1, and an edge state is supported. This shows that the `limiting to different groupings' argument frequently used in the discussion of the SSH model and the Kitaev chain (see Refs.~\cite{Batra,Ma2022Dec} and Fig.~\ref{fig:trimer}) is not so easily generalisable to more complicated systems. 

This is not the only topological edge state of the SSH4 model. The $T$ matrix also supports edge states when $\bar\mu=\pm\sqrt{\gamma_{13}^2+\gamma_2^2}$. In this case the eigenvalue is $\gamma_2/\gamma_4$, and it looks like the edge state of an SSH chain with the $\gamma_1=\gamma_3$ couplings ignored. Our analysis using the $T$-matrix thus shows that there are two quantities which are independently responsible for the creation and destruction of topological edge states. Namely, $\gamma_{13}^2/(\gamma_2\gamma_4)$ and $\gamma_2/\gamma_4$.

\begin{figure}[h]
    \includegraphics[width=\linewidth]{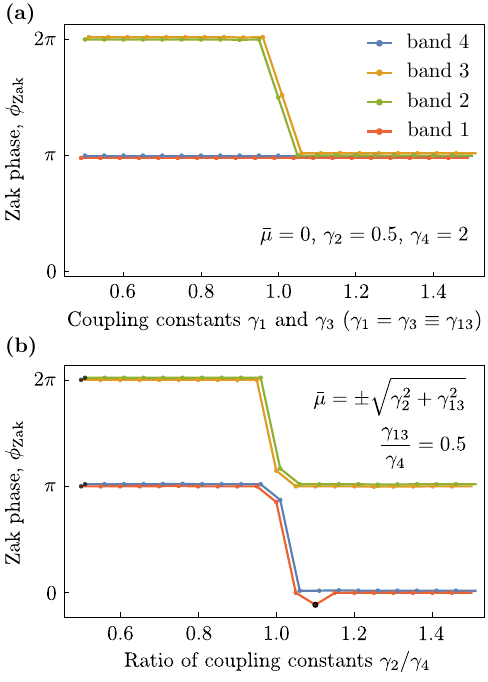}
    \caption{Zak phase of the SSH4 model with $\gamma_1=\gamma_3$. Both plots show the Zak phase $\phi_{\text{Zak}}$ across the two transitions. The bands are numbered from lowest to highest in energy. A small diagonal $\pm0.01$ shift is added to the Zak phases for clarity purposes. \textbf{(a)} The Zak phase across the $\gamma_{13}^2 = \gamma_2\gamma_4$ transition point. \textbf{(b)} The Zak phase across the $\gamma_2=\gamma_4$ transition for the edge states. The dip at $\gamma_2/\gamma_4=1.1$ is a numerical artifact. }
    \label{fig:ssh4zak}
\end{figure}

We will now verify that our $T$-matrix results are in agreement with the conventional method of computing Bloch-band topological invariants. We compute the Zak phase of each band using \eqref{eqn:zakphaseeigvec}. The results, summarised in Fig.~\ref{fig:ssh4zak}, show that the Zak phase is quantised to multiples of $\pi$. This quantisation of the Zak phase is evidence that the restricted SSH4 model is indeed topologically nontrivial. We note that there are two families of edge states which can be created or destroyed independently of the other. The first family consists of the edge state between\footnote{As in figure \ref{fig:ssh4zak} we number the bands from lowest to highest in energy.} bands 2 and 3, and the second family consists of the edge states between bands 1 and 2, and between bands 3 and 4. Each family can only support 1 or 0 edge states. As such, the topological phases of the SSH4 model can be classified by the group is $\mathbb{Z}_2 \times \mathbb{Z}_2$.

So far we have only analysed the SSH4 model protected by parity symmetry. In the dimer case, both parity symmetry and chiral symmetry \cite[Sec.~II.C]{Chiu2016Aug} impose the same requirement $v_1=v_2$ in  Eq.~\eqref{eqn:ssh}, and both are individually responsible for the topological protection of the edge mode \cite{lan}. However, in the tetramer SSH4 case chiral symmetry imposes a different condition $v_i=0$ than parity symmetry, see Eq.~\eqref{eqn:ssh4}. For a brief analysis, on the transfer matrix level, chiral symmetry is satisfied as long as
\begin{equation}
\sigma_z T(E) \sigma_z = T(-E).
\end{equation}
This relation provides protection only to the edge mode between the second and third bands at $E=0$. The other edge modes protected by parity symmetry are no longer protected when only chiral symmetry is present.

Finally, our analysis of the SSH4 model shows that unlike the Bloch Hamiltonian, the transfer matrix lends itself easily to expanding the size of the unit cell. This comes at the expense of making long-range couplings more difficult to introduce. However, this also means that all SSH$n$ models and continuous systems can still be described by $2\times 2$ transfer matrices, and as such is susceptible to our analysis thus far. Our method therefore generalises to any real one-dimensional quantum mechanical or photonic systems, without discretisation.

\subsection{Dimerised Photonic Crystal}

The transfer matrix is a well-known tool in analysing photonic crystals, and have previously been used to prove a bulk-edge correspondence in the interface of two inversion symmetric one-dimensional photonic crystals \cite{Xiao2014Apr}. We note that our transfer matrix approach is especially useful for photonic systems since the analogous Hamiltonian for photonic systems is more difficult to work with than the transfer matrix \cite{lan}.

\begin{figure}[b]
    \includegraphics[width=\linewidth]{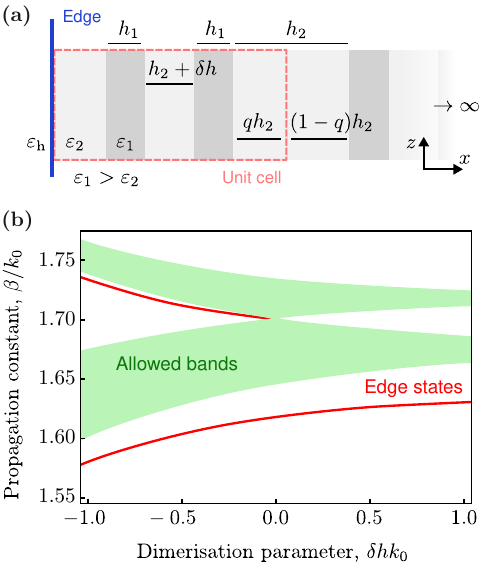}
    \caption{ 
      \textbf{(a)} Dimerised dielectric photonic crystal inspired by the SSH model. We have assumed for simplicity that the material is nonmagnetic. %
      The unit cell is chosen so that a fraction $1-q$ of the low index layer is at the left edge. For $q = 0.5$ the unit cell is symmetric.
      \textbf{(b)} Numerical $T$-matrix calculation of the band structure of the asymmetric DPC with $q=1$. The edge states shown as red lines. The dimensionless parameters used are as follows: $\varepsilon_1=4,\varepsilon_2=2,k_0h_1=k_0h_2=2.$ The host material parameters has been chosen to be $\varepsilon_h=\mu_h=1$. We see that there is an edge state below the two bands, which is not predicted by the SSH model. In addition, the edge state in between the two bands appears in the wrong side $\delta h<0$ compared to the SSH model. Since the unit cell is asymmetric (i.e. Eq.~\eqref{eqn:tinversion} is violated), the edge states are not topologically protected, and the corresponding plot in $\mathrm{SL}(2,\mathbb{R})$ space is going to be 3 dimensional.}
    \label{fig:dpcbandasym}
\end{figure}

The transfer matrix for a single unit cell of a photonic crystal in the full wave basis $(E_\perp,H_\perp)$ is given by \cite{orfanidis}
\begin{align}\label{eqn:tlambda}
\hat{T}_\Lambda = \prod_{i=N}^1 \begin{pmatrix}
\cos(\phi_i) & \iu Z_i^{-1} \sin(\phi_i)\\
\iu Z_i\sin(\phi_i) & \cos(\phi_i)
\end{pmatrix},
\end{align}
where $\phi_i = k_{x,i}h_i$, $Z_i^{TE} = Z_0\mu_i k_0/k_{x,i}$ in the TE case, and $Z_i^{TM} = Z_0k_{x,i}/(\varepsilon_i k_0)$ in the TM case. Here, $Z_0$ is the impedance of free space, $k_0 = \sqrt{\varepsilon_i\mu_i}\omega/c$ is the total wavenumber, so that $k_{x,i} = \sqrt{\varepsilon_i\mu_i k_0^2-\beta^2}$ where $\beta$ is the wavenumber in the continuous-translation-invariant direction ($z$ direction in Fig.~\ref{fig:dpcbandasym}~(a)). The role of this transfer matrix is to take in the electroamgnetic waves at a certain position $x$ and calculate the electromagnetic waves one unit-cell away $x+\Lambda$. Namely, $\hat{T}_\Lambda: (E_\perp(x),H_\perp(x)) \mapsto (E_\perp(x+\Lambda),H_\perp(x+\Lambda))$. Here, $\Lambda=\sum_i h_i$ is the unit cell length.

\begin{figure*}
    \includegraphics[width=0.90\linewidth]{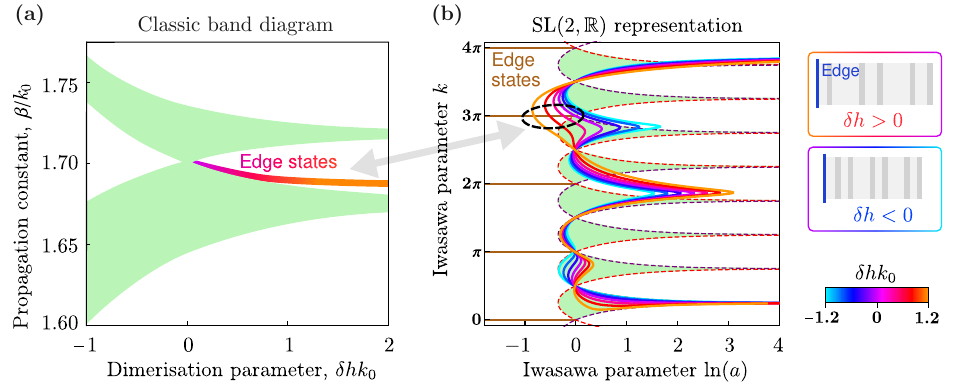}
    \caption{ The band structure of the symmetric, $q=1/2$, dimerised photonic crystal embedded in $\mathrm{SL}(2,\mathbb{R})$ and projected to the $n$ given by Eq.~\eqref{eqn:tinversion} in the TM case. Each curve represents a family of $T$-matrices $T_\Lambda(\beta)$ for a different value of $\delta h$. All other parameters are identical to Fig.~\ref{fig:dpcbandasym}~(b). The bands are ordered so that $\beta$ increases as the curve goes from bottom to top. When the curve touches a black line, a topologically protected edge state is supported. The green regions are the allowed bands: Each point on the curve inside the green region corresponds to a bulk propagating mode. }
    \label{fig:dpckan}
\end{figure*}

We consider a dimerised photonic crystal (DPC) as illustrated in Fig.~\ref{fig:dpcbandasym}, with alternating lengths for the low index layer. We will analyse the DPC with the framework we have built so far.

We first note that any eigenmode of the DPC must satisfy the impedance matching condition at the boundary
\begin{equation}\label{eqn:impedancematching}
    \frac{E_{\perp}}{H_{\perp}} = Z_h = \begin{cases}
Z_0\dfrac{ \mu_h k_0}{-\iu \kappa} & \quad \text{(TE)} \vspace{0.3cm} \\ 
Z_0 \dfrac{-\iu \kappa}{\varepsilon_h k_0} & \quad  \text{(TM)}
\end{cases}
\end{equation}
where $\kappa = \sqrt{\beta^2-\varepsilon_h\mu_h k_0^2}$ and $\varepsilon_h,\mu_h$ are the relative permittivity and permeability of the host material~\cite{toftul}.

Then, by changing basis from $(E_\perp,H_\perp)$ to $(F_-,F_+)$:
\begin{align}
F_{\pm}=E_\perp\pm Z_h H_\perp
\end{align}
the impedance matching condition \eqref{eqn:impedancematching} on the left and right edges become $F_-=0$ and $F_+=0$ respectively. As such, the condition for an edge mode becomes identical to the one we obtained for the SSH models:
\begin{align}\label{eqn:dpctedge}
\bar{T}_{\Lambda}(\beta) \begin{pmatrix}
0\\1
\end{pmatrix} = \alpha \begin{pmatrix}
0\\1
\end{pmatrix}\comma \alpha \in (-1,1).
\end{align}
We also note than, when we restrict to the values of $\beta$ so that the field decays outside the structure, $Z_h$ is imaginary, and since the off diagonal components of each multiplicand in \eqref{eqn:tlambda} is always imaginary, the $T$ matrix in the $F_{\pm}$ basis is always real. As such, the DPC is susceptible to the same analysis as we have done for the real SSH models.

\begin{figure}
    \centering
    \includegraphics[width=\linewidth]{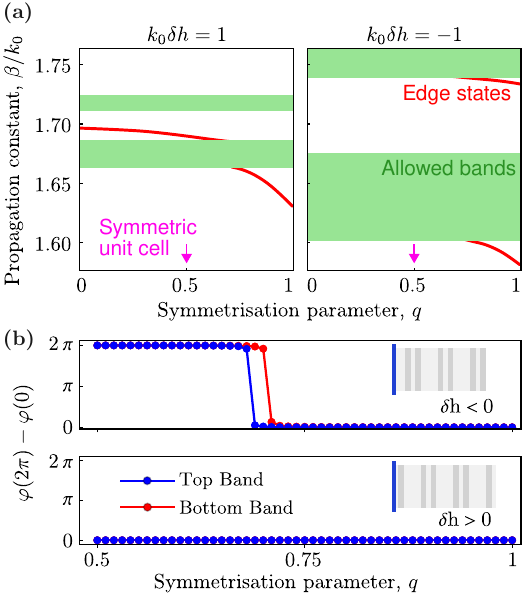}
    \caption{
    \textbf{(a)} Numerically computed edge states of the DPC for $k_0  \delta h =\pm 1$, and various symmetrisation parameters $q$. We see that in both cases the appearance of extended edge state at $q \approx 0.7$ is accompanied by the edge state between the bands switching from $\delta h >0$ to $\delta h<0$.
    \textbf{(b)} The winding of the relative phase of the entries in the $T$ eigenvector, $\varphi(2\pi)-\varphi(0)$, across the transition. The transition in the winding number is only allowed when an edge state is either created or destroyed at the boundary of an allowed band. When $\delta h<0$, we see that there is a delay between the creation of the edge state at the top and bottom bands. When $\delta h>0$, no delay was detected.
    }
    \label{fig:dpcband_and_phi}
\end{figure}

We may compute the band structure of TM waves in the DPC with $q=1$ using the $T$-matrix numerically, with the bulk bands computed using Bloch's theorem, and the edge states by numerically finding solutions to \eqref{eqn:dpctedge} in the following manner: The parameter space spanned by $\beta$ and $\delta h$ is discretised to a grid of size $ 500\times 50$ respectively. A cell is then marked as an edge state if \eqref{eqn:dpctedge} is satisfied within an adjustable tolerance. This results Fig.~\ref{fig:dpcbandasym}~(b) and we see that the results are unlike the SSH model. The edge state appears on the wrong side $\delta h<0$, which in the coupled mode approximation (Appendix \ref{app:coupledmode}) means $\gamma_1 > \gamma_2$. In addition, we see an extended edge state appearing below the two bands, which spans through both regions $\delta h<0$ and $\delta h>0$. %

The condition that $T$ is parity invariant, Eq.~\eqref{eqn:tinversion}, is exactly the condition that the unit cell is symmetric $q=1/2$. In this case, all edge states are topological and the $\mathrm{SL}(2,\mathbb{R})$ trajectories for the TM case are illustrated in Fig.~\ref{fig:dpckan}. The TE case gives similar results, so we will omit it for brevity. From this figure, we see that edge states are supported if and only if $\delta h >0$, in agreement with the coupled mode approximation in Appendix \ref{app:coupledmode}. There seems to be no extended edge states in this case. As a result, we see that the extended edge mode in Fig.~\ref{fig:dpcbandasym}~(b) is a result of the defect introduced by the asymmetric prescription of the unit cell, and is not a property of the bulk.

However, this extended edge mode might still be tied to our SSH-like edge mode somehow. We may plot the edge states as a function of $q$ in exactly the same manner as we did in Fig.~\ref{fig:dpcbandasym}~(b). The results are plotted in Fig.~\ref{fig:dpcband_and_phi}~(a) and we see that the appearance of the extended edge state is always accompanied by the SSH-like edge state changing sides from $\delta h >0$ to $\delta h<0$. To investigate this further, we may plot the the relative phase between the entries in the Bloch eigenvector. Since this phase can only change when an edge state touches one of the bands, it provides a way to accurately detect the precise $q$ at which edge states are created or destroyed. This is plotted in Fig.~\ref{fig:dpcband_and_phi}~(b) and we see that, when $\delta h<0$, there seems to be a small delay between the creations of the SSH-like and extended edge state, indicating that the two edge modes may not be very intimately connected after all. But it is still suspicious that the two transitions appear so close together, and that this delay was not detected for the $\delta h>0$ case, so this \qqoute{coincidence} may be worth future investigation.

We have characterised the edge states of the DPC for the allowed values of $\beta$ such that the field decays away from the periodic structure. The extended edge state is understood to arise from the defect introduced by the asymmetric perturbation $q\neq 0$, and its existence seems to be weakly tied to the topological edge state between the two SSH-like bands predicted by the coupled mode approximation (Appendix \ref{app:coupledmode}).

\section{Conclusion}

We have developed an alternative approach to analyse the topological properties of one-dimensional topological systems by employing the transfer matrix. Using this technique, we have obtained a simple proof for the Zak phase bulk-edge correspondence and have characterised the edge states of the generalized tetramer SSH model, the so-called SSH4 model. We have demonstrated that the SSH4 model supports topological edge states when the parity symmetry condition~\eqref{eqn:tinversion} is imposed, and that the behaviour of the SSH4 model is not entirely intuitive: Unlike the dimer SSH model, an edge state may appear even if the chain is cut at the weakest bond. Our analytical results have been verified by numerical computation of the Zak phases of each band. Finally, we have employed our approach of the transfer matrix to study dimerized photonic crystals, and we have demonstrated that a requirement for the edge states to be topological is for the unit cell to be symmetric with $q=1/2$. The existence of the extended edge mode has also been observed to be tied to the cut, $\delta h <0$ or $\delta h>0$, where the SSH-like edge state resides. 

\begin{acknowledgements}
Y.K. thanks Alexander Khanikaev, Alexey Slobozhanyuk, and Dmitry Zhirikhin for useful comments.  This work was supported by the Australian National University and the Australian Research Council (Grants No. DP200101168 and No. DP210101292).
\end{acknowledgements}

\appendix 

\section{Proof of Zak Phase-Eigenvector Phase relation}\label{app:zakphaseeigvec}

In this appendix section, we will show that the Zak phase of a $PT$-symmetric system may be computed by considering the ratio of the first and last entry of the Bloch eigenvector. We will take the time reversal symmetry to be complex conjugation. In this way, our result will hold not only for the SSH$n$ models with real hoppings considered in the paper, but also for example, the complex-hermitian SSH$n$ models. The case of anti-parity-time symmetric non-Hermitian Su-Schrieffer-Heeger (SSH) models is discussed in Ref.~\cite{Wu2021PRB}.

Consider a system with $N$ sites per unit cell. Without loss of generality, we may assume that $N$ is even, since for a system with odd $N$, we can double the number of sites by merging 2 unit cells. 
We label the amplitude in the $m^\text{th}$ site as $a_m$. 
In general, due to $PT$ symmetry, the Bloch eigenvector will satisfy
\begin{align}
	\ket{K} \propto \Sigma_x \ket{K}^*
\end{align}
where $\Sigma_x$ is the permutation matrix with $1$ on the anti-diagonal, and zero elsewhere.
This means, we may always write the Bloch eigenvector in the form
\begin{align}\label{aeqn:blocheigvec}
	\ket{K} =\frac{\eu^{\iu \phi(K)/2}}{\sqrt{1+r_2^2(K)+ r_3^2(K) + \dots}}\begin{pmatrix}
   \eu^{-\iu \phi(K)/2}\\r_2(K)\eu^{-\iu \phi_2(K)}\\r_3(K)\eu^{-\iu \phi_3(K)}\\\vdots\\r_3(K)\eu^{\iu \phi_3(K)}\\r_2(K)\eu^{\iu \phi_2(K)}\\\eu^{\iu \phi(K)/2}
    \end{pmatrix}.
\end{align}
Here, $r_i$ and $\phi_i$ are real-valued functions of $K$.
Note that the choice of reciprocal gauge $\eu^{\iu \phi(K)/2}$ is physical: It is required for the Zak phase to have an unambiguous interpretation of mean displacement. To prove this, we first fix the first unit cell to have position $0$. As a result, the Bloch eigenvector is related to the real-space Hamiltonian eigenstate as
\begin{align}\label{aeqn:dfteigvec}
	\ket{K,\text{Real}} = (1,e^{\iu K},e^{2\iu K},\cdots) \otimes \ket{K}
\end{align}
As a result, for all $K$, we have that
\begin{align}
	\braket{1}{K,\text{Real}} =a_1 = \text{const.}
\end{align}
Here, $\ket{1}$ is the position basis vector corresponding to the leftmost site, $K=2\pi m/L$ is the crystal momentum, $L$ is the number of unit cells, and $m$ is the unit cell number.
We may fix the global phase to $0$, so that the Bloch eigenvector is given by \eqref{aeqn:blocheigvec}. We thus see that allowing the phase of the first entry in $\ket{K}$ to vary is equivalent to choosing a different, translated coordinate system for each value of the crystal momentum $K$.
	
We are now ripe to prove the assertion \eqref{eqn:zakphaseeigvec} in the main text. Defining $\ket{K,\text{Sym}} =\eu^{-\iu \phi(K)/2} \ket{K}$ to be the RHS of \eqref{aeqn:blocheigvec} without the global phase factor, the action of the gauge transformation on the derivative is
\begin{align}
	\bra{K,\text{Sym}}\eu^{-\iu \phi(K)/2} (-\iu)\partial_K  \eu^{\iu \phi(K)/2}\ket{K,\text{Sym}}\nonumber\\=\bra{K,\text{Sym}}\left[-\iu \partial_K +\frac{\phi'(K)}{2}\right]\ket{K,\text{Sym}}.
\end{align}
While computing the inner product $\bra{K,\text{Sym}}\partial_K\ket{K,\text{Sym}}$, the $m^\text{th}$ and $(N-m)^\text{th}$ terms always cancel, so we are left with
\[
	\phi_\text{Zak} = \int \limits_0^{2\pi} \frac{\phi'(K)}{2} \dd K = \frac{\phi(2\pi)-\phi(0)}{2}
\]
as required.
	
The generalisation to continuous systems is straightforward by taking $N \to \infty$ with a symmetric finite difference scheme.

Recently, it has been brought to our attention that \cite{Alexandradinata2014Apr} has previously obtained a general expression for the Zak phase in terms of the eigenvalue of the inversion operator acting on $\ket{K=0}$ and $\ket{K=\pi}$, which can be used to derive Eq.~\eqref{eqn:zakphaseeigvec} from the main text more straightforwardly.

\section{Four Branches of Solutions for $T$}\label{app:foursolutions}

In this appendix section, we clarify the four branches of solutions mentioned in Sec. \ref{sec:proofzakphase}. From parity symmetry, we have the constraint given in \eqref{eqn:tinversion}. Substituting this back into $T$ and taking the trace gives us
\begin{align}
    \mathrm{Tr}(T)=\sec (k) \left(a \cos (2 k)+\frac{1}{a}\right)
\end{align}
Solving for the condition $|\mathrm{Tr}(T)|=2$ then, we obtain four different solutions
\begin{align}
    a=\frac{1}{\pm \cos(k)\pm \sin(k)} \quad \text{or} \quad  \frac{1}{\pm \cos(k)\mp \sin(k)}.
    \label{eq:a_sol}
\end{align}
Since $a>0$ we are required to choose only the positive solutions in Eq.~\eqref{eq:a_sol} for each $k$. Substituting \eqref{eq:a_sol} back into $T$ gives us the four branches mentioned in the main text. Note that the conditions $\det(T)=1$ and $|\operatorname{Tr}(T)| = |2\cos K|=2$ fix all eigenvalues to have the same sign, and be $\pm 1$. This way, $K$ is quantized to be 
$K \in \pi\mathbb{Z}$. In the case $k\not\in\pi\mathbb{Z}$ these solutions will not be diagonalisable and the Bloch state can be recovered from the unique $T$-eigenvector. The results are summarised in Table \ref{tab:Tfoursolutions}.

\bgroup
\begin{table}[]
\centering
\caption{Table summarising the four solutions of $T \in\mathrm{SL}(2;\mathbb{R})$ satisfying both \eqref{eqn:tinversion} and $|\mathrm{Tr}(T)|=2$. The value of $K$ is obtained as the argument of the eigenvalue of $T$. The value of $\varphi$ is the phase difference between the two entries of the unique eigenvector of $T$.}
\label{tab:Tfoursolutions}
\begin{ruledtabular}
\begin{tabular}{ccc}
        $T$ & $K$ & $\varphi$  \\ 
        \hline 
   $ \begin{pmatrix}
 -\frac{\cos (k)}{\sin (k)+\cos (k)} & \frac{\sin (k)}{\sin (k)+\cos (k)} \\
 -\frac{\sin (k)}{\sin (k)+\cos (k)} & -\frac{2 \sin (k)+\cos (k)}{\sin (k)+\cos (k)}
\end{pmatrix}$ & $\pi $& $ \pi$  \\ [0.7cm]
$\left(
\begin{array}{cc}
 \frac{\cos (k)}{\cos (k)-\sin (k)} & -\frac{\sin (k)}{\cos (k)-\sin (k)} \\
 \frac{\sin (k)}{\cos (k)-\sin (k)} & \frac{\cos (k)-2 \sin (k)}{\cos (k)-\sin (k)} 
\end{array}
\right)$ & 0 & 0 \\ [0.7cm]
$\left(
\begin{array}{cc}
 \frac{\cos (k)}{\sin (k)-\cos (k)} & \frac{\sin (k)}{\cos (k)-\sin (k)} \\
 \frac{\sin (k)}{\sin (k)-\cos (k)} & -\frac{\cos (k)-2 \sin (k)}{\cos (k)-\sin (k)} 
\end{array}
\right)$ & $\pi$ & 0 \\ [0.7cm]
$\left(
\begin{array}{cc}
 \frac{\cos (k)}{\sin (k)+\cos (k)} & -\frac{\sin (k)}{\sin (k)+\cos (k)} \\
 \frac{\sin (k)}{\sin (k)+\cos (k)} & \frac{2 \sin (k)+\cos (k)}{\sin (k)+\cos (k)} 
\end{array}
\right)$ & 0 & $\pi$ 
\end{tabular}
\end{ruledtabular}
\end{table}
\egroup
	
\section{Coupled Mode Approximation to the Dimerised Photonic Crystal}\label{app:coupledmode}
In the coupled mode approximation, we model each high index material as a dielectric waveguide. The interactions between neighbouring waveguides can then be modelled using perturbation theory. The full treatment is provided in \cite{Okamoto2006}, but the physical intuition is rather simple. Each waveguide supports a mode which can be thought of being localised around the waveguide. As multiple waveguides are brought closer together, their modes will start to couple, and this coupling must be symmetric. This is illustrated in the figure \ref{fig:sshpc}. 

\begin{figure}[h]
	\tikzset{squiggly/.style={decorate, decoration={snake, amplitude=0.5mm}}}
	\begin{tikzpicture}[scale=0.7]
	\fill[lightgray] (-2,-2) rectangle (-1,2);
	\fill[lightgray] (2,-2) rectangle (1,2);
	\fill[lightgray] (5,-2) rectangle (6,2);
	\fill[lightgray!10] (-1,-2) rectangle (1,2);
	\fill[lightgray!10] (2,-2) rectangle (5,2);
	\draw[thick] (-2,2)--(-2,-2);
	\draw[thick] (-1,2)--(-1,-2);
	
	\draw[thick] (2,2)--(2,-2);
	\draw[thick] (1,2)--(1,-2);
	\draw[thick] (5,2)--(5,-2);
	\draw[thick] (6,2)--(6,-2);
	
	\draw[<->,thick, squiggly] (-1,-1)--(1,-1) node[midway, below]{$\kappa_1$};
	\draw[<->,thick, squiggly] (2,-1)--(5,-1) node[midway, below]{$\kappa_2$};
	
	\draw[|-|,thick] (-2,-2.2)--(-1.05,-2.2) node[midway,below]{$h_1$};
	\draw[|-|,thick] (-0.95,-2.2)--(0.95,-2.2) node[midway,below]{$h_2+\delta h$};
	\draw[|-|,thick] (1.05,-2.2)--(1.95,-2.2) node[midway,below]{$h_1$};
	\draw[|-|,thick] (2.05,-2.2)--(4.95,-2.2) node[midway,below]{$h_2$};
	\draw[|-|,thick] (5.05,-2.2)--(6,-2.2) node[midway,below]{$h_1$};
	
	\draw[|-|,thick] (-2,2.2)--(5,2.2) node[midway,above]{unit cell};
	
	\node[] () at (-1.5,0) {$\varepsilon_1$};
	\node[] () at (1.5,0) {$\varepsilon_1$};
	\node[] () at (-1.5,-1) {$\phi_1$};
	\node[] () at (1.5,-1) {$\phi_2$};
	\node[] () at (5.5,-1) {$\phi_3$};
	\node[] () at (0,0) {$\varepsilon_2$};
	\node[] () at (3.5,0) {$\varepsilon_2$};
	\node[] () at (5.5,0) {$\varepsilon_1$};
	
	\node[rectangle, draw=black, fill=white, inner sep=2pt] () at (7,-1) {$\varepsilon_1>\varepsilon_2$};
	
	\end{tikzpicture}
	
	\caption{Dimerised dielectric photonic crystal inspired by the SSH model. We have assumed for simplicity that the material is nonmagnetic. In the coupled mode approximation, each high index layer (dark) is modelled as a waveguide and the low index layer serves only for separation. Each $\phi_i$ represent the amplitude of the electromagnetic wave in each mode, and $\kappa_i$ denote the coupling strengths between each neighbouring waveguide.}
	\label{fig:sshpc}

\end{figure}

If we consider only nearest neighbour interactions, the only possible form of the equations of motion are:

\begin{alignat}{3}
\iu \frac{\dd \phi_1}{\dd z} &=\beta_0 \phi_1 &+\kappa_1 \phi_2 & \nonumber\\
\iu \frac{\dd \phi_2}{\dd z} &=\kappa_1 \phi_1 &+\beta_0 \phi_2 &+\kappa_2 \phi_3 \nonumber\\
\iu \frac{\dd \phi_3}{\dd z} &= & \kappa_2 \phi_2 &+\beta_0 \phi_3 
\end{alignat}
Here, $z$ is the direction of continuous translation symmetric, $\beta_0$ is the bare propagation constant of each lone waveguide, and $\kappa_i$ are the coupling coefficients. These equations look identical to the SSH model, and one might think that the DPC should indeed reproduce the SSH model in some way. It turns out though, that this correspondence does not hold for all the allowed modes in the DPC. For modes with higher wavenumber in the $z$ direction $\beta$, the field will not have enough momentum to be a travelling wave in the low index material. As a result, the field looks like a superposition of guided modes around each high index material. However, when the propagation constant $\beta$ is chosen to be sufficiently small, the waves will have enough momentum to take the form of travelling waves in the low index material. As such, our underlying assumption for the coupled mode approximation breaks down. This is the reason for the appearance of extended bulk modes in the DPC.

\bibliography{refs}     %

\end{document}